\documentstyle[prl,aps,multicol,epsf]{revtex}
\begin{document}
\draft
\title{Dynamics of spin ladders}
\author{R. Eder}
\address{Institut f\"ur Theoretische Physik, Universit\"at W\"urzburg,
Am Hubland,  97074 W\"urzburg, Germany
}
\date{\today}
\maketitle
\begin{abstract}
We derive an approximate theory for Heisenberg spin ladders with two legs
by mapping the spin dynamics onto the problem of hard-core `bond-Bosons'. 
The parameters of the Bosonic Hamiltonian are obtained by matching anomalous 
Green's functions to Lanczos results and we find evidence for a strong 
renormalization due to quantum fluctuations. Various dynamical spin correlation
functions are calculated and found to be in good agreement with Lanczos 
results. We then enlarge the effective Hamiltonian to describe the coupling 
of the Bosonic spin fluctuations to a single hole injected into the system 
and treat the hole-dynamics within the `rainbow-diagram' approximation by 
Schmidt-Rink {\em et. al.} Theoretical predictions for the single hole 
spectral function are obtained and found to be in good agreement with
Lanczos results.
\end{abstract}
\begin{multicols}{2}
\section{Introduction}
Copper-oxides with a $CuO_2$ plane containing line-defects, which results in
ladder-like arrangments of $Cu$-atoms\cite{Hiroi}, 
have recently received considerable 
attention\cite{DagottoRice}. In addition to being interesting physical 
systems themselves\cite{Azuma,Uehara}, 
they may be considered, on the theoretical side, as an 
important stepping-stone to understand the fully $2D$ $CuO_2$ planes
of cuprate superconductors. Namely their special geometry makes
two-legged ladders a realization of a particularly simple `RVB'-type
state, and developing a `technology' for handling such
RVB states clearly is one of the key issues in the theory of high-temperature
superconductors. In the following, we want to derive
a simple theoretical description of the spin dynamics and single-hole
dynamics of ladders. Thereby we follow an approach which is 
similar in spirit to the Landau theory of Fermi liquids: we use a simple 
picture of an `RVB spin liquid' and its excitation spectrum and
parameterize the dispersion of the spin excitations by few parameters,
which are then obtained by matching the results to
Lanczos calculations and finite-size analysis. Using this 
parameterization we make quantitative predictions
for various dynamical correlation functions
which can be compared to exact diagonalization (and experiment). 
The theory actually allows to make rather detailed
predictions, which in all cases studied are in good agreement 
with Lanczos results.
We also discuss possibilities to extend the calculation to
$2D$ systems.\\
We consider the standard $t-J$ model on a $2$-leg ladder
with $N$ rungs. More precisely, the Hamiltonian reads
\begin{equation}
H = -\sum_{\langle i,j\rangle} (t_{ij}\;
\hat{c}_{i,\sigma}^\dagger \hat{c}_{j,\sigma} + H.c.)
+\sum_{\langle i,j\rangle} J_{ij} \; \bbox{S}_i \cdot
\bbox{S}_j.
\end{equation}
Here $\langle i,j\rangle$ denotes a summation over all pairs of
nearest neighbors on the ladder, for bonds along the legs
we choose
$t_{ij}=t$ and $J_{ij}=J$, for bonds along
the rungs $t_{ij}=t_\perp$ and $J_{ij}=J_\perp$. 
In the following we adopt the values $J_\perp=J$, $t_\perp=t$,
and $J/t=0.5$, which may be roughly appropriate for
actual the materials. We choose the $x$-axis along the direction of the
legs, the $y$-axis along the rungs.
\section{Spin dynamics}
\begin{figure}
\epsfxsize=8cm
\vspace{-2.0cm}
\hspace{0.0cm}\epsffile{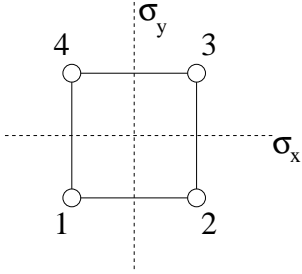}
\vspace{-3.0cm}
\narrowtext
\caption[]{Labelling of the sites in the $2\times 2$
plaquette.}
\label{fig1} 
\end{figure}

\begin{figure}
\epsfxsize=10cm
\vspace{-0.0cm}
\hspace{-0.5cm}\epsffile{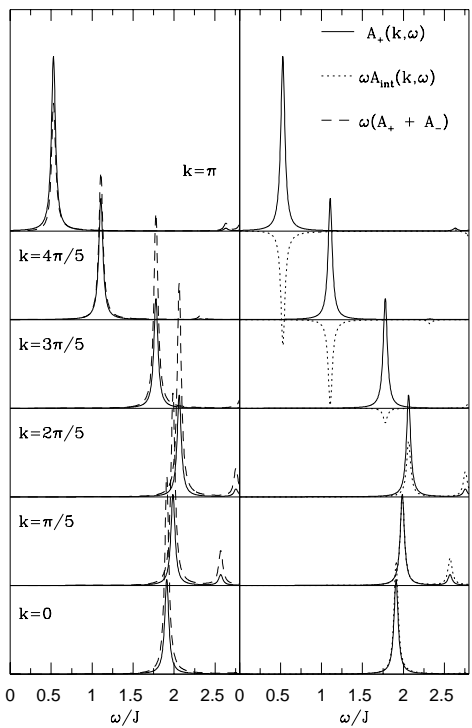}
\vspace{-0.0cm}
\narrowtext
\caption[]{Spectral functions $A_+(k,\omega)$ (full line),
$\omega( A_+(k,\omega) + A_-(k,\omega))$ (dashed line) and
$\omega A_{int}(k,\omega)$ (dotted line) obtained by
Lanczos diagonalization of a $2\times 10$ ladder. 
$\delta$-functions are replaced by Lorentzians of width $0.1J$.}
\label{fig2} 
\end{figure}

\begin{figure}
\epsfxsize=10cm
\vspace{-0.0cm}
\hspace{-0.5cm}\epsffile{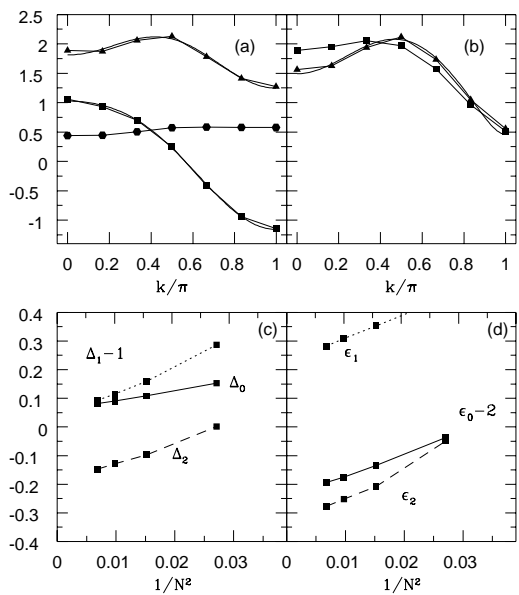}
\vspace{-2.0cm}
\narrowtext
\caption[]{(a): $\tilde{\epsilon}_k$ (triangles),
$\tilde{\Delta}_k$ (squares), and $Z$ (hexagons) as obtained from the
pole-strengths of the Greens functions plotted versus
momentum. The continuous line
gives Fourier expansions with the lowest $3$ harmonics.\\
(b): $\omega_k$ calculated from (\ref{tdisp})
(triangles) compared to the exact excitation energies
(squares).\\
(c) and (d): Scaling of the Fourier coefficients
of $\Delta$ and $\epsilon$ with the length of the ladder.}
\label{fig3} 
\end{figure}

\begin{figure}
\epsfxsize=10cm
\vspace{-0.0cm}
\hspace{-0.5cm}\epsffile{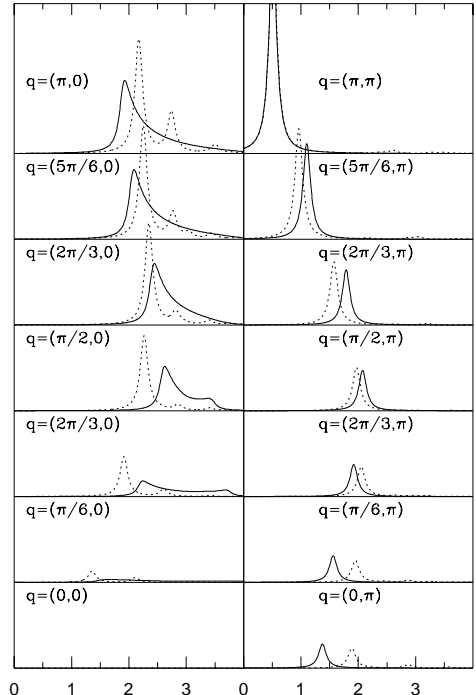}
\vspace{-0.5cm}
\narrowtext
\caption[]{Spin correlation function $S(\bbox{q},\omega-i\cdot 0.1J)$
obtained by numerical evaluation
of (\ref{scs} (full line) compared to the result
of Lanczos diagonalization on an $N$$=$$12$ ladder (dotted line).
Spectra for $q_y$$=$$0$ are multiplied by
a factor of $5$.}
\label{fig4} 
\end{figure}

To begin with, we consider the
case of half-filling and, following Ref.\cite{Gopalan} 
define the following operators:
\begin{eqnarray}
s_{ij}^\dagger &=& \frac{1}{\sqrt{2}}
(\hat{c}_{i,\uparrow}^\dagger \hat{c}_{j,\downarrow}^\dagger
- \hat{c}_{i,\downarrow}^\dagger \hat{c}_{j,\uparrow}^\dagger),
\nonumber \\
t_{ij,x}^\dagger &=& 
\frac{-1}{\sqrt{2}}(
\hat{c}_{i,\uparrow}^\dagger \hat{c}_{j,\uparrow}^\dagger
-\hat{c}_{i,\downarrow}^\dagger \hat{c}_{j,\downarrow}^\dagger)
\nonumber \\
t_{ij,y}^\dagger &=& 
\frac{i}{\sqrt{2}}(
\hat{c}_{i,\uparrow}^\dagger \hat{c}_{j,\uparrow}^\dagger +
\hat{c}_{i,\downarrow}^\dagger \hat{c}_{j,\downarrow}^\dagger),
\nonumber \\
t_{ij,z}^\dagger &=& \frac{1}{\sqrt{2}}
(\hat{c}_{i,\uparrow}^\dagger \hat{c}_{j,\downarrow}^\dagger
+ \hat{c}_{i,\downarrow}^\dagger \hat{c}_{j,\uparrow}^\dagger).
\label{triplets}
\end{eqnarray}
These create either a singlet or the three components of a triplet
on the sites $i$ and $j$. 
Following a large number of
workers\cite{Gopalan,Endres,Sierra,Martins}, 
we will start out from the `rung-RVB state',
$|\Psi_0\rangle = \prod_{n=0}^{N}
s_{n,n+\hat{y}}^\dagger |0\rangle$,
which is the ground state in the limit
$J_\perp/J\rightarrow \infty$
(here  $i+\hat{y}$ denotes the
nearest neighbor of $i$ in $y$-direction).
Let us now consider the effect of switching on $J$. Obviously this
will create `fluctuations' in the vacuum state, and it is of
importance to clarify the nature of these. Due to the
product nature of the vacuum it is sufficient to
consider just one $4$-site plaquette, see Figure \ref{fig1}
for the labelling of sites. For $J$$=$$0$ the ground state is
$|0\rangle= s_{14}^\dagger s_{23}^\dagger |vac\rangle$,
with energy $-(3/2)J_\perp$. It might appear that the 
fluctuation with the lowest cost in energy would be
a `$90$-degree swap', i.e. a transition to a state with singlets
along the legs: $|1\rangle$$=$$s_{12}^\dagger s_{43}^\dagger |vac\rangle$.
This state has energy $-(3/2)J$, i.e. it is
degenerate with the vacuum in the limit
$J_\perp$$=$$J$. This line of thinking, however, is incorrect.
The reason is that $|1\rangle$ is not orthogonal to the
vacuum, more precisely one finds $\langle 1|0\rangle = 1/2$.
The most natural way to proceed is
to form the orthogonal
complement $|2\rangle = |1\rangle - (1/2) |0\rangle$,
which (after normalization) is found to be
\[
|2\rangle = \frac{1}{\sqrt{3}}
(t_{14,x}^\dagger t_{23,x}^\dagger + t_{14,y}^\dagger t_{23,y}^\dagger
+ t_{14,z}^\dagger t_{23,z}^\dagger) |vac\rangle.
\]
The (manifestly rotation invariant)
expression on the r.h.s. thereby is nothing but the Clebsch-Gordan
combination of two triplets along the rungs into a singlet.\
\noindent
This shows, that the `true' excitation is the double triplet,
rather than a state with singlets along the legs, or, put another
way, that the state $|1\rangle$ is redundant and can be discarded
if we keep $|0\rangle$ and $|2\rangle$. The same holds true
for the last candidate state for a fluctuation,
namely two triplets along the legs coupled to a singlet:
\[
|3\rangle = \frac{1}{\sqrt{3}}
(t_{12,x}^\dagger t_{43,x}^\dagger + t_{12,y}^\dagger t_{43,y}^\dagger
+ t_{12,z}^\dagger t_{43,z}^\dagger) |vac\rangle.
\]
Again we find that this state
is not orthogonal to the vacuum either, and
orthogonalization again yields $|2\rangle$.
The only fluctuation we have to take into account therefore
is the formation of two triplets on adjacent rungs,
the respective matrix element is $\langle t_{14,z} t_{23,z} |H|
s_{14}^\dagger s_{23}^\dagger\rangle = J/2$. The energy increases
by $2J_\perp$, which we interpret as twice the `energy of formation'
of a single triplet.
Let us now assume that a double triplet has been
formed, and consider its further development.
First, the two triplets can `recombine' and
we return to the vacuum;  the matrix element for this
process again is $J/2$. Second,
the two triplets can `swap their species'. More precisely, for 
$\alpha$$\neq$$\beta$ there is a matrix element of the form
$\langle t_{14,\beta} t_{23,\beta}|H|
t_{14,\alpha}^\dagger t_{23,\alpha}^\dagger \rangle$$=$$-J/2$.
The last possibility is that  
one of the triplets interacts with a singlet on the as yet
`untouched' rung next to it. To study this in more detail we
consider a $4$-site plaquette containing spin $1$.
It is convenient to form the states
\[
|\pm \rangle =
\frac{1}{\sqrt{2}}
( t_{14,z}^\dagger s_{23}^\dagger \pm s_{14}^\dagger t_{23,z}^\dagger)
|vac\rangle
\]
which do have definite parities under the mirror operations $\sigma_x$ 
and $\sigma_y$ (see Figure \ref{fig1}). They are triplets with $z$-component 
$0$.
Next we consider which state could mix with one of these states.
A first possibility would be two `rung triplets', coupled
to a triplet:
\[
|4\rangle = \frac{1}{\sqrt{2}}
(t_{14,x}^\dagger t_{23,y}^\dagger - t_{14,y}^\dagger t_{23,x}^\dagger)
|vac\rangle.
\]
However, this state has positive parity under $x$-reflection,
whereas $|\pm \rangle$ both have negative parity. 
This state therefore
cannot be admixed\cite{Gopalan}. The next possibility is
\[
|5 \rangle =
\frac{1}{\sqrt{2}}
( t_{12,z}^\dagger s_{43}^\dagger - s_{12}^\dagger t_{43,z}^\dagger)
|vac\rangle,
\]
(which does have the required negative $x$-parity), but 
this state is actually identical to $|-\rangle$.
Finally, one could think of two triplets along the legs coupled to
form a triplet:
\[
|6\rangle =\frac{1}{\sqrt{2}}
(t_{12,x}^\dagger t_{43,y}^\dagger - t_{12,y}^\dagger t_{43,x}^\dagger)
|vac\rangle.
\]
but, in fact, $|6\rangle=|+\rangle$.
The only process possible for the rung-triplet
therefore is to `propagate', i.e. exchange its position with
a singlet on a neighboring rung. The
hopping element is $(\langle +|H|+\rangle -
\langle-|H-\rangle)/2 = J/2$. 
The `90-degree swap'  to a triplet
along the leg (see the state $|5\rangle$) would give a redundant state and,
most importantly, `anharmonic processes' whereby a 
triplet decays into two
triplets (see the states $|5\rangle$ or $|6\rangle$) 
are not possible either. Clearly,
this absence of anharmonicity leads to a considerable
simplification of the physics - it is the ultimate reason why,
as will be seen later, the spin correlation function in a ladder is
remarkably `coherent'.\\
There is one last process we need to discuss.
It may happen that two triplets of unlike species
which have been created in different pair creation processes
`collide'. More precisely, for $\alpha \neq \beta$
there is a matrix element of the type
$\langle  t_{14,\beta} t_{23,\alpha} | H|
 t_{14,\alpha}^\dagger t_{23,\beta}^\dagger \rangle$$=$$J/2$,
which describes two triplets of unlike species
`hopping over' one another.\\
We now represent the presence of a triplet on the $n^{th}$ rung
by the presence of a `book-keeping boson'
created by $t_{n,\alpha}^\dagger$. 
Grouping the three possible triplet states into
a single 3-vector, $\bbox{t}_n$$=$$(t_{n,x}, t_{n,y}, t_{n,z})$,
we can describe all
processes involving non-redundant 
states by the following 
manifestly rotation-invariant Hamiltonian
\begin{eqnarray}
H &=& J_\perp \sum_{ n }
\bbox{t}_{n}^\dagger \cdot \bbox{t}_{n}^{}
\nonumber \\
&+&\frac{J}{2} \sum_{ n }
(\;\bbox{t}_{n}^\dagger \cdot \bbox{t}_{n+1}^\dagger + H.c.\;)
+\frac{J}{2} \sum_{ n }
(\; \bbox{t}_{n}^\dagger \cdot \bbox{t}_{n+1}^{} + H.c.\;)
\nonumber \\
&-&\frac{J}{2}
\sum_n (\;
\bbox{t}_n^\dagger \cdot \bbox{t}_{n+1}^\dagger\;
\bbox{t}_{n+1}^{} \cdot \bbox{t}_{n}^{}
-\bbox{t}_n^\dagger \cdot \bbox{t}_{n+1}^{}\;
\bbox{t}_{n+1}^\dagger \cdot \bbox{t}_{n}^{}\;).
\label{eff}
\end{eqnarray}
This Hamiltonian has also been derived by Gopalan
{\em et al.}\cite{Gopalan}.
The form of the Hamiltonian (\ref{eff})
suggests a quite obvious approximation, namely
to break down the quartic terms in a BCS-like fashion, e.g.:
\begin{eqnarray}
t_{n,\beta}^\dagger t_{n+1,\beta}^\dagger
t_{n+1,\alpha}^{} t_{n,\alpha}^{}
&\rightarrow& \Delta\; t_{n,\beta}^\dagger t_{n+1,\beta}^\dagger
+ \Delta^*\; t_{n+1,\alpha}^{} t_{n,\alpha}^{}
\nonumber \\
\Delta &=& \langle t_{n+1,\alpha}^{} t_{n,\alpha}^{} \rangle.
\end{eqnarray}
As a next step, we might assume that the hopping integrals
and pair creation matrix elements in (\ref{eff}) are renormalized in a
Gutzwiller-like fashion, to mimic the effect
of the hard-core constraint. After Fourier transform,
we would thus arrive at an approximate Hamiltonian of the form
\begin{eqnarray}
H_{eff} &=& \sum_{k>0} \;[\;(
\tilde{\epsilon}_k \;  \bbox{t}_k^\dagger \cdot \bbox{t}_k^{}
+ \tilde{\epsilon}_{-k} \;  \bbox{t}_{-k}^\dagger \cdot \bbox{t}_{-k}^{})
\nonumber \\
&\;& \;\;\;\;\;\;\;\;+
(\tilde{\Delta}_k \; \bbox{t}_k^\dagger \cdot \bbox{t}_{-k}^\dagger + H.c.)
\;].
\label{eff1}
\end{eqnarray}
Assuming that the $t$ in this
Hamiltonian are {\em free Bosons} this is readily
solved by the ansatz
\begin{eqnarray}
\bbox{\gamma}_{k,\alpha}^{} &=& u_k \bbox{t}_k^{} + v_k 
\bbox{t}_{-k}^\dagger
\nonumber \\
\bbox{\gamma}_{-k,\alpha}^\dagger &=& 
v_k \bbox{t}_k^{} + u_k \bbox{t}_{-k}^\dagger,
\end{eqnarray}
to give the ($3$-fold degenerate) dispersion
\begin{equation}
\omega_k = \sqrt{ \tilde{\epsilon}_k^2 - \tilde{\Delta}_k^2}.
\label{tdisp}
\end{equation}
The parameters $\tilde{\epsilon}_k$, $\tilde{\Delta}_k$ should be calculated
in some approximate fashion, by mean-field and Gutzwiller-approximation
to the original Hamiltonian (\ref{eff})\cite{Gopalan}. 
We note, however, that there are a number of difficulties
 with such an approach:
the density of Bosons $n_b$ can be obtained from the
spin correlation function along the rungs using the
identity $-3(1-n_b)/4 + n_b/4= \langle \bbox{S}_i \cdot  \bbox{S}_{i+\hat{y}}
\rangle$. The spin correlation function thereby can be estimated
from exact diagonalization results.
For $J_\perp/J=1$ we thus obtain $n_b\approx0.3$.
The quantum fluctuations thus are strong, and we may expect
that the matrix elements for propagation and pair creation
in (\ref{eff}) are heavily renormalized. On the other hand, the
matrix elements of the quartic terms in (\ref{eff}) will not
be renormalized at all due to the hard-core constraint, because
these terms do not change the
Boson occupation of any site. We may thus expect a subtle interplay
of Gutzwiller projection and mean-field decomposition and 
to avoid poorly controlled
approximations we resort to numerical techniques to
extract the parameters of the Hamiltonian (\ref{eff1}) and,
in doing so,
moreover check the quality of the mapping to noninteracting Bosons
{\em per se}. To that end we define the operator
\[
\tau_{n}^\dagger = \frac{1}{2}
( n_{n,\uparrow} n_{n+\hat{y},\downarrow}
- n_{n,\downarrow} n_{n+\hat{y},\uparrow}
- S_n^+ S_{n+\hat{y}}^-
+ S_n^- S_{n+\hat{y}}^+).
\]
If the rung $(n,n+\hat{y})$ is in a singlet state,
$\tau_{n}^\dagger$ transforms it into the $z$-component of the
triplet; if the rung is in any of the triplet states, the state is
annihilated. This operator, while acting entirely
within the Hilbert space of the original ladder system,
thus may be viewed as a realization
of the hard-core Boson creation operator. 
Next, using standard Lanczos techniques, we evaluate the
following Green's functions:
\begin{eqnarray}
A_+(k,\omega) &=& \Im \frac{1}{\pi}
\langle\Psi_0| \tau_{-k} \frac{1}{\omega - (H-E_0)-i0^+}
\tau_{-k}^\dagger |\Psi_0\rangle,
\nonumber \\
A_-(k,\omega) &=& \Im \frac{1}{\pi}
\langle\Psi_0| \tau_{k}^\dagger \frac{1}{\omega - (H-E_0)-i0^+}
\tau_k |\Psi_0\rangle,
\nonumber \\
A_{int}(k,\omega) &=& \Im \frac{1}{\pi}
\langle\Psi_0| \tau_{k}^\dagger \frac{1}{\omega - (H-E_0)-i0^+}
\tau_{-k}^\dagger |\Psi_0\rangle,
\end{eqnarray}
where $|\Psi_0\rangle$ ($E_0$) denote the ground state wave function
(energy) of the ladder and $\tau_k^\dagger$ is the
(1 dimensional) Fourier transform of $\tau_n^\dagger$. 
Our goal is to map the spin excitations of the ladder
onto a system of free `Quasi-Bosons' goverened by the Hamiltonian
(\ref{eff1}). If we want to compare spectral properties, we have to
take into account that the {\em spectral weight} of
the hard-core Bosons may be strongly renormalized.
For example,  as a rigorous identity we have
$\langle [\tau_i, \tau_i^\dagger]\rangle$$=$$1-4n_b/3$$\approx$$0.6$, 
rather than
$1$ as it would be for free Bosons. To discuss the
spectral functions we therefore assume that
$\tau_k$$\rightarrow$$\sqrt{Z} t_k$, where $t_k$ is a free Boson operator,
and $Z$ the wave function renormalization constant.\\
Assuming then that the free Bosons $t_k$ are indeed described
by a Hamiltonian of the form (\ref{eff1}) one can derive
the following expressions for the above Green's functions:
\begin{eqnarray}
\omega A_{int}(k,\omega) &=& -\frac{Z\tilde{\Delta}_k}{2} 
\delta(\omega - \omega_k)
\nonumber \\
\omega( A_+(k,\omega) + A_-(k,\omega))
&=& Z\tilde{\epsilon}_k \delta(\omega - \omega_k).
\nonumber \\
 A_+(k,\omega) - A_-(k,\omega)
&=& Z \delta(\omega - \omega_k).
\label{theory}
\end{eqnarray}
In other words, the wave function renormalization $Z$
as well as the energy $\tilde{\epsilon}_k$ and pair creation
amplitude $\tilde{\Delta}_k$ can be read off from the
dispersion of the {\em pole strength}
in the Green's functions (\ref{theory}). Then, using the obtained
values of $\tilde{\epsilon}_k$ and  $\tilde{\Delta}_k$
to calculate the dispersion of the {\em excitation energy}
from (\ref{tdisp}) and comparing with the actual numerical values
should provide a stringent cross-check for the validity
of the mapping to the free-Boson Hamiltonian (\ref{eff1}).
Figure \ref{fig2} compares $A_+(k,\omega)$ and 
$\omega A_{int}(k,\omega)$ and $\omega( A_+(k,\omega) + A_-(k,\omega))$.
To begin with, the $A_+(k,\omega)$ shows a series of sharp dispersive peaks,
which however trace out a rather unusual dispersion with
a shallow local minimum at $k$$=$$0$. Next, $\omega A_{int}(k,\omega)$
has a pronounced low energy peak too, which coincides with that
of $A_+(k,\omega)$. The residuum changes sign at approximately
$k=\pi/2$ (note that $A_{int}$ is an {\em off-diagonal} matrix element of the
resolvent operator; its residuum therefore need not be positive definite),
the magnitude is nearly identical at $k$$=$$0$ and $k$$=$$\pi$.
This suggests a $k$-dependence of the form $\tilde{\Delta}_k\propto \cos(k)$,
as one would expect for a pair amplitude corresponding to
pair creation on nearest neighbors. Similarly,
$\omega( A_+(k,\omega) + A_-(k,\omega))$ shows low energy peaks
coniciding with those of $A_+(k,\omega)$. The dispersion of the
weight, however, is more complicated than for the pairing amplitude.
\noindent
For a quantitative
analysis, we proceed to Figure \ref{fig3}. 
Part (a) shows first of all the weight $Z$ obtained from
$A_+(k,\omega) - A_-(k,\omega)$. Remarkably enough,
$Z$ is quite independent of $k$ and moreover close to the
estimate of $0.6$ obtained from
$\langle [\tau_i, \tau_i^\dagger]\rangle$. The fact that
$Z$ is indeed fairly $k$-independent is a first indication
for the applicability of a free Boson Hamiltonian.
To avoid a proliferation of adjustable parameters
we take the $k$-average, which is $Z$$=$$0.53$ (nearly
independent of $N$ for $N>6$). Using this value of $Z$ we
then calculated the
$\tilde{\epsilon}_k$ and $\tilde{\Delta}_k$ and
Fourier transform them with respect to $k$.
For simplicity, we terminate the Fourier series after the
third (i.e. $\cos(2k)$-like) term. This gives already a very satisfactory
fit to the data, as seen in Figure \ref{fig3}.
Figure \ref{fig3}b shows the comparison of the
numerical excitation energy and the dispersion calculated from
(\ref{tdisp}), using the values of $\tilde{\epsilon}_k$ and $\tilde{\Delta}_k$
extracted from the weights of the correlation functions.
There is quite reasonable agreement, which indicates that the
effective free-Boson Hamiltonian itself is a quite good description
of the spin dynamics. \\
We proceed to extrapolate the results to the infinite
chain. To that end, we performed the
calculation for different $N$ and
calculated the Fourier coefficients of $\tilde{\epsilon}_k$ and
$\tilde{\Delta}_k$. Then, Figures \ref{fig3}(c) and (d) show plots of these
Fourier coefficients vs. $1/N^2$. The plots rather obviously suggest
that all of the Fourier coefficients can be
\noindent
expressed as $c(N)=c_\infty + c'/N^2$ to good accuracy.
The extrapolated values $c_{\infty}$ are
given in Table \ref{tab1}. 
We note in passing that these
values give a `spin gap' (i.e. the energy of the triplet
with $k=\pi$) of $0.51J$, in agreement with exact 
diagonalization\cite{Barnes} and DMRG calculations\cite{White}.
Let us now briefly discuss these values.
For the pairing amplitude $\tilde{\Delta}_k$ we have a by far dominant
$\cos(k)$-component, $\Delta_1$, consistent with pair 
creation/annihilation on 
nearest neighbors. The second-nearest neighbor amplitude is substantially
smaller, and also the uniform component is very small.\\
The results for the renormalized energy $\tilde{\epsilon}_k$ are
more surprising. To begin with, the constant term is
$1.77J$, rather than $J_\perp$$=$$J$, as one would expect.
A possible explanation for this strong increase is that a
triplet on (say) the rung number $i$ blocks the
`pair creation' of triplets on the pairs $(i-1,i)$ and
$(i,i+1)$. The corresponding loss of fluctuation energy
increases the energy cost for creating
a triplet, i.e. the on-site energy of a triplet. In addition,
presence of a triplet will partially inhibit the motion
of other triplets and thus cause a further loss of
delocalization energy. Quite obviously these
effects combined are quite substantial.
Next, the coefficient of the nearest-neighbor harmonic
is rather small (only $0.21$). This may be understood simply
in terms of a Gutzwiller-like reduction of the mobility due to
the `excluded volume' occupied by other triplets. Moreover,
the terms which describe the `exchange hopping' and
effectively propagate a triplet by one lattice site
may interfere destructively with the `ordinary'
motion of a triplet, and thus reduce the effective hopping integral.
A further surprise is the large 
amplitude for second-nearest neighbor hopping.
Here one could envisage processes where the propagating triplet
encounters a pair of triplets, recombines with one of them and
thereby transfers its momentum to the remaining triplet,
so that the propagating triplet effectively has been 
transferred by two sites. Such a process would be proportional to the
probability of finding a quantum fluctuation, which is rather high
in the present case.
All in all, the data show that the propagation of the
triplets is strongly renormalized in the relatively dense and
strongly interacting `background' of the other triplets.\\
Assuming that the `effective Hamiltonian' (\ref{eff1}) with the
extrapolated parameters in Table \ref{tab1} gives a good
description of the spin dynamics, 
we proceed to calculate dynamical correlation functions relevant to
experiment. We consider the spin operator on a single
rung and `translate it' into the language of the hard-core
Bosons. The operator $S_\alpha(k_\perp$$=$$\pi)$ turns a singlet
into an $\alpha$-triplet and vice versa; the other two types
of triplets are annihilated, whence:
$\bbox{S}(k_\perp$$=$$\pi)\rightarrow\bbox{t}^\dagger$$+$$\bbox{t}$.
Next, the operator $S_z(k_\perp$$=$$0)$ annihilates a
singlet and converts e.g. $t_x^\dagger$$\rightarrow$$i t_y^\dagger$
(compare (\ref{triplets})).
$\bbox{S}(k_\perp$$=$$0)$ therefore
must be bilinear in the triplet operators and the only
possibility is $\bbox{t}^\dagger \times \bbox{t}$.
From its acting on $t_x^\dagger$ the prefactor must be
$-i$, whence:  $\bbox{S}(k_\perp$$=$$0)$$\rightarrow$$-i\bbox{t}^\dagger 
\times \bbox{t}$. Obviously this is also the only way to construct
a Hermitean operator ($\bbox{S}(k_\perp$$=$$0)$ is the operator of
total spin on one rung) from the triplets. 
A subtle point is the renormalization of the spectral weight:
whereas $\bbox{S}(k_\perp$$=$$\pi)$ actually changes the
number of Quasi-Bosons and thus
should be renormalized in a similar fashion as the
hard-core Boson addition and removal spectra,
$\bbox{S}(k_\perp$$=$$0)$ does
not change the Boson occupation of any rung and thus should
remain essentially unrenormalized. Based on these considerations, we
multiply $\bbox{S}(k_\perp$$=$$\pi)$ by $\sqrt{Z}$, but leave
$\bbox{S}(k_\perp$$=$$0)$ unrenormalized. Then, upon 
Fourier transformation of the
single rung operators and switching to the bond-Bosons
we obtain the spin correlation functions:
\begin{eqnarray}
S(q,\pi,z)&=& \frac{Z}{2} \frac{(u_q - v_q)^2}{z - \omega_k},
\nonumber \\
S(q,0,z) &=& \frac{1}{2N}
\sum_k \frac{ (u_{k+q} v_k-u_{-k} v_{-(k+q)})^2}
{z - (\omega_{k+q}+ \omega_k)}.
\label{scs}
\end{eqnarray}
Note that $S(q,0,z)\rightarrow 0$ for $q\rightarrow 0$; this is what
must come out because unlike conventional spin-wave theory 
the `rung-RVB' state is an exact singlet, and both Hamiltonians,
(\ref{eff}) and (\ref{eff1}) are
rotationally invariant; the ground state thus is an
exact singlet, whence the operator of total
spin must annihilate it.
The obtained spin correlation functions are shown
in Figure \ref{fig4}.
We first note the rather different intensity of the
two correlations functions. As 
\noindent
could be expected
on the basis of (\ref{scs}), the correlation function for
$q_y$$=$$\pi$ consists essentially of a single peak, whereas
the spectrum for $q_y$$=$$0$ has more `cusp-like' appearance.
It is interesting to note that these
differences are reproduced quite well by the results of
Lanczos calculation, which are also shown in Figure (\ref{fig4}):
the $q_y$$=$$\pi$ spectrum is indeed remarkably sharp,
with nearly all the spectral weight
concentrated in just one peak for every momentum.
It should be noted that the agreement of the dispersion
is not really surprising, this was actually seen already
in Figure \ref{fig3}. The dispersion of the {\em peak intensity},
however, also agrees very well with the Lanczos result and thus
provides further evidence for the correctness of the
mapping to the Boson Hamiltonian.\\
By contrast, the spectra for $q_y=0$ usually consist
of several peaks and one can already envisage how these spectra
develop into the cusp-like spectra produced by
our theory. The dispersion of the
`peaks' for $q_y$$=$$0$ is in good agreement with theory,
showing a shallow maximum at $q_x$$\approx$$3\pi/5$ and approaching
zero for $q_x\rightarrow 0$.
All in all, the spin correlation function is obviously very
well reproduced by our calculation.
We proceed to a discussion of the `energy correlation function'
$e_{\alpha}(\bbox{q},\omega)$,
i.e. the dynamical correlation function of the operator
\[
H_{\alpha}(\bbox{q}) = \frac{1}{\sqrt{N}}
\sum_i e^{i \bbox{q} \cdot \bbox{R}_i }
\bbox{S}_i \cdot \bbox{S}_{i+\hat{\alpha}}.
\]
Apart from being a further probe of the spin dynamics, this
correlation function plays a key role in the
theory of phonon-assisted two-magnon absorption
by Lorenzana and Sawatzky\cite{jose}. It therefore
can be probed experimentally in infra-red
absorption measurements. Here we restrict ourselves to
the operator $H_{y}(\bbox{q})$, which is easy
to `translate' to the Quasi-Boson system. Namely one can replace
\[
\bbox{S}_i \cdot \bbox{S}_{i+\hat{\alpha}} \rightarrow
\bbox{t}_i^\dagger \cdot \bbox{t}_{i}^{}
\]
whence
\[
e_y(\bbox{q},z) =
 \frac{3}{2N}
\sum_k \frac{ (u_{k+q} v_k+u_{-k} v_{-(k+q)})^2}
{z - (\omega_{k+q}+ \omega_k)}.
\]
Since $H_{y}(\bbox{q})$ does not change the number of
Bosons on any rung, we have not added a factor of $Z$.
The other correlation function, $e_x(\bbox{q},\omega)$
is more difficult. In principle, the exchange along a given
bond along the legs of the ladder is nothing but the
respective part of the effective Hamiltonian (\ref{eff}).
However, as has been discussed above, if we go over
to the simplified Hamiltonian (\ref{eff1}) there is 
quite a strong renormalization and also a generation of
additional next-nearest-neighbor hopping terms. We believe this is
difficult to treat in a reasonably controlled 
\begin{figure}
\epsfxsize=10cm
\vspace{-0.0cm}
\hspace{-0.5cm}\epsffile{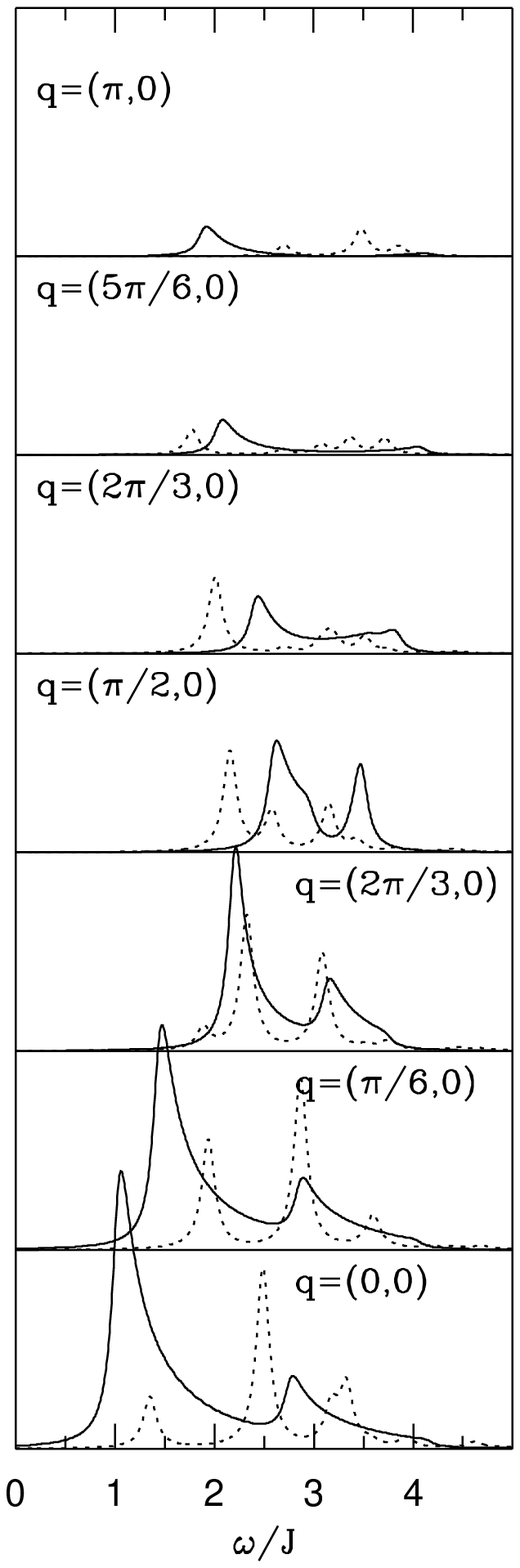}
\vspace{-0.5cm}
\narrowtext
\caption[]{Energy correlation function $e(\bbox{q},\omega-i\cdot 0.1J)$
obtained by numerical evaluation
of (\ref{scs} (dotted line) compared to the result
of Lanczos diagonalization on a $N=12$ ladder (full line).
Note that $q_y=\pi$ cannot be meaningfully defined
and is therefore omitted.}
\label{fig5} 
\end{figure}
\noindent
way and therefore
omit a discussion of this correlation function.
Then, $e_y(\bbox{q},\omega)$ is shown in Figure \ref{fig5}
and compared to Lanczos results.
Unlike the spin correlation function, the detailed agreement of the 
lineshape between the theory and Lanczos is not very good, with
the discrepancy being particularly large around $q$$=$$0$. There, the
Lanczos result shows an intense peak at an energy of
$\approx 2.5J$, which is
completely absent in the theoretical spectra. We believe that
this peak corresponds to a `bi-triplet', i.e. a bound state
of two triplets on nearest neighbors. Two triplets on nearest
neighbors can efficiently lower their energy by using
the exchange-like interaction described by the 
quartic terms in (\ref{eff}).
This may produce a virtual bound state similar as the
bimagnon excitation familiar from the Raman spectrum of the
$2D$ Heisenberg antiferromagnet. 
Our simple particle-hole
picture for the correlation function
naturally does not take into account such
a bound state formation and therefore cannot
reproduce this resonance. On the other hand,
for momenta which are more distant
from $q$$=$$0$ the agreement between Lanczos and theory
improves somewhat. The theoretical spectra
reproduce the lower edge and
`width' of the spetra reasonably well, and also the
$\vec{q}$-dependence of the total spectral weight is
approximately correct.
\section{Hole dynamics and photoemission}
\begin{figure}
\epsfxsize=7cm
\vspace{-0.5cm}
\hspace{0cm}\epsffile{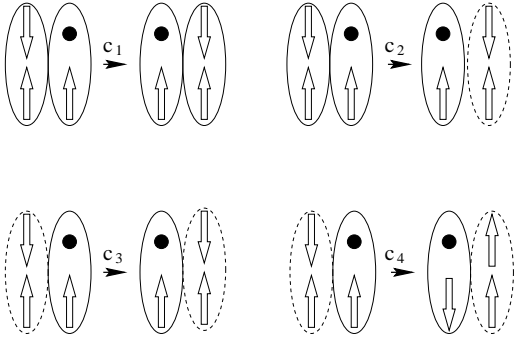}
\vspace{-0.0cm}
\narrowtext
\caption[]{Possible interactions between a hole and 
a doubly occupied rung. Each process is labeled by
the prefactor of the terms in the Hamiltonian (\ref{c_hopp}) 
which describes it. The possible processes are:
exchange with a singlet ($c_1$), exchange with a singlet
which is transformed into a triplet ($c_2$), exchange with
a triplet whereby the triplet remains unchanged ($c_3$),
exchange with a triplet whereby the triplet
changes its $S_z$ ($c_4$).}
\label{processes} 
\end{figure}

\begin{figure}
\epsfxsize=10cm
\vspace{-0.0cm}
\hspace{-0.5cm}\epsffile{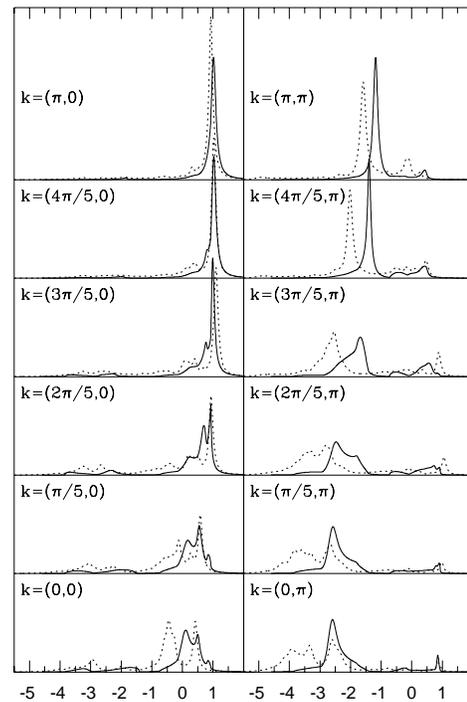}
\vspace{-0.5cm}
\narrowtext
\caption[]{Hole spectral function by self-consistent
Born aproximation (full line) compared to 
the spectrum of $\tilde{c}_{\bbox{k},\sigma}$
obtained by Lanczos diagonalization in a
$2\times 10$ ladder (dotted line). The ratio $J/t$$=$$0.5$.}
\label{fig7} 
\end{figure}

Next, we proceed to a theory for photoemission\cite{Haas}.
To that end, we need to study the
motion of a single hole in the `spin background'. The previous discussion
has shown that we needed to retain only singlets and triplets
along the rungs and that, as far as the spin dynamics is concerned,
all other possible states are redundant. This suggests to
discuss the added hole in such a way that it `fits' with the
rung basis used for discussing the spin dynamics. 
We define bonding and antibonding states of one electron along
a rung:
\begin{equation}
a_{n,k_y,\sigma}^\dagger = \frac{1}{\sqrt{2}}
( \hat{c}_{n,\sigma}^\dagger + e^{ik_y} \hat{c}_{n+\hat{y},\sigma}^\dagger ),
\end{equation}
where we have represented the parity under $\sigma_x$
as momentum $0$ or $\pi$ in $y$-direction.
These states have an `on site energy' of $-\cos(k_y)t_\perp$ and we
now consider their propagation.
The goal thereby is to interpret a singly occupied rung
as being occupied by an `effective Fermion', which has the
$z$-spin and $k_y$ of the respective single-electron state.
Then we want to enlarge the effective Hamiltonian
(\ref{eff1}) by terms describing the propagation
of these effective Fermions as well as their interaction with the
triplet excitations.\\
The form of the terms in this effective Hamiltonian can be 
inferred already from the requirements of positive parity
under $\sigma_x$, spin rotation and time-reversal invariance
and Hermiticity. We note that the rung-singlet and
$a_{n,0,\sigma}^\dagger$ have positive
parity under $\sigma_x$,
whereas the triplets and $a_{n,\pi,\sigma}^\dagger$
have negative parity.
Then, from the spinors $a$ we can construct two $3$-vectors
\begin{eqnarray}
\bbox{S}_{n,m} &=& \sum_{k_y}
a_{m,k_y,\sigma}^\dagger \;\vec{\tau}_{\sigma,\sigma'} \;
a_{n,k_y+\pi,\sigma'} ,
\nonumber \\
\bar{\bbox{S}}_{n,m} &=& \sum_{k_y}
a_{m,k_y,\sigma}^\dagger \; \vec{\tau}_{\sigma,\sigma'} \;
a_{n,k_y,\sigma'} ,
\end{eqnarray}
where $\vec{\tau}$ is the vector of Pauli matrices. Thereby
$\bbox{S}_{n,m}$ has odd parity under exchange of the
legs, $\bar{\bbox{S}}_{n,m}$ has even parity, and
we have $\bbox{S}_{n,m}^\dagger$$=$$\bbox{S}_{m,n}$. 
Both vectors are odd under time reversal (as is $\bbox{t}$).
The processes we need  to describe are shown schematically
in Figure \ref{processes}: by virtue
\noindent
the hopping term a singly occupied rung can exchange
its position with a `fully occupied' rung. Thereby
the fully occupied rung can remain in a singlet,
change from singlet to triplet (and vice versa), 
remain in a given triplet state, or remain in a triplet state
but change its $S_z$.
These processes can be described
by the following coupling terms, whose form follows
simply from the requirements of spin rotation
invariance, even parity, and Hermiticity:
\begin{eqnarray}
H&=& c_1 \sum_{n,k_y}
[\; \bar{n}_n
a_{n,k_y,\sigma}^\dagger
 a_{n-1,k_y,\sigma} + H.c.\;]
\nonumber \\
&+& c_2 \sum_n [\;
\bbox{t}_n^\dagger\cdot (
\bar{n}_{n+1} \bbox{S}_{n,n+1} + \bar{n}_{n-1} \bbox{S}_{n,n-1})
 + H.c.]
\nonumber \\
&+& c_3 \sum_n [\;
\bbox{t}_{n-1}^\dagger \cdot \bbox{t}_{n}\;
\sum_{k_y,\sigma} a_{n,k_y,\sigma}^\dagger a_{n-1,k_y,\sigma}
+ H.c.\;]
\nonumber \\
&+& c_4 \sum_n [\;
i \bar{\bbox{S}}_{n,n-1} \cdot( \bbox{t}_{n}^\dagger \times \bbox{t}_{n-1})
+ H.c. \;],
\label{c_hopp}
\end{eqnarray}
where
\[
\bar{n}_n = 1-\bbox{t}_n^\dagger \cdot \bbox{t}_n.
\]
To determine the numerical values of the coefficients $c_i$ 
we start with the state
$|1\rangle= a_{1,k_y,\uparrow}^\dagger s_{2}^\dagger|vac\rangle$.
Denoting the hopping term along the legs by $H_t$ we find:
\[
H_t |1\rangle =\frac{t}{2}(
a_{2,k_y,\uparrow}^\dagger s_{1}^\dagger
+ a_{2,k_y+\pi,\uparrow}^\dagger t_{1,z}^\dagger
- \sqrt{2} a_{2,k_y+\pi,\downarrow}^\dagger t_{1,+}^\dagger)|vac\rangle.
\]
Next, starting from the state 
$|1'\rangle= a_{1,k_y,\uparrow}^\dagger t_{2,z}^\dagger|vac\rangle$
we obtain
\[
H_t |1'\rangle =\frac{t}{2}(
a_{2,k_y+\pi,\uparrow}^\dagger s_{1}^\dagger
+ a_{2,k_y,\uparrow}^\dagger t_{1,z}^\dagger
+ \sqrt{2} a_{2,k_y,\downarrow}^\dagger t_{1,+}^\dagger)|vac\rangle.
\]
Using these two equations we find
$c_1$$=$$t/2$, $c_2$$=$$t$, $c_3$$=$$t/2$, and $c_4$$=$$-t$.\\
Next, we consider the action of the exchange terms
along the legs. Namely the electron on a singly occupied rung
can exchange with an electron belonging to a singlet or triplet
on a neighboring rung.
Again we can write down the general form of the
Hamiltonian as
\begin{eqnarray}
H&=& 
c_2' \sum_n [\;
( \bbox{t}_{n+1}^\dagger +\bbox{t}_{n-1}^\dagger )\cdot 
\bbox{S}_{n,n}  + H.c.]
\nonumber \\
&+& c_4' \sum_n \;
i \bar{\bbox{S}}_{n,n} \cdot( \bbox{t}_{n\pm1}^\dagger 
\times \bbox{t}_{n\pm1} ).
\end{eqnarray}
Denoting the exchange term along the legs as $H_J$, we find
\begin{eqnarray}
H_J |1\rangle &=&\frac{J}{4}(
 a_{1,k_y+\pi,\uparrow}^\dagger t_{2,z}^\dagger
- \sqrt{2} a_{1,k_y+\pi,\downarrow}^\dagger t_{2,+}^\dagger)|vac\rangle,
\nonumber \\
H_J |1'\rangle &=&\frac{J}{4}(
 a_{1,k_y+\pi,\uparrow}^\dagger s_{2}^\dagger
+ \sqrt{2} a_{1,k_y,\downarrow}^\dagger t_{2,+}^\dagger)|vac\rangle,
\end{eqnarray}
whence $c_2'$$=$$J/2$ and $c_4'$$=$$-J/2$.
Together with the Hamiltonian for the triplet-dynamics,
Eq. (\ref{eff1})
then gives a complete description for the
hole motion in the ladder.\\
Having found all possible ways of interaction between
the hole and the spin excitations, we need to consider
a suitable approximation to handle these terms.
To begin with, by treating the terms $\propto c_1, c_2$
in (\ref{c_hopp}) in a mean-field like way, we may hope to
obtain a `renormalized hopping' integral for the hole.
Namely we replace $\bar{n}_n
\rightarrow (1- n_b)$, with $n_b=0.3$ the density of
triplets (see above).
Next, in the term $\propto c_3$ we replace 
\begin{eqnarray}
\bbox{t}_{n-1}^\dagger \cdot \bbox{t}_{n}
&\rightarrow& \langle \bbox{t}_{n-1}^\dagger \cdot \bbox{t}_{n} \rangle
\nonumber \\
&=& \frac{3}{N} \sum_k \cos(k)\; v_k^2.
\end{eqnarray}
We thus obtain the `effective hopping integral'
\[
t_{eff} = \frac{t}{2}[\;(1-n_b) + \frac{3}{N} 
\sum_k \cos(k)\;v_k^2\;].
\]
Numerical evaluation shows that this is a very small
quantity, $t_{eff}$$=$$0.27t$, the reason being that the
second term on the r.h.s. is negative and of quite
appreciable magnitude.\\
The form of the terms $\propto c_2, c_2'$
suggests the `rainbow diagram' approximation 
due to Schmidt-Rink {\em et al.}\cite{SchmidtRink} to 
treat them. 
This still leaves us with the terms $\propto c_4, c_4'$;
detailed investigation shows, however, that
the matrix elements of these terms are proportional to
higher powers of the (small) coherence factor $v_k$;
we take this as a justification to neglect these terms.\\
Next, we briefly discuss additional `renormalizations' which might occur.
In the same way as a triplet on
some given rung will `block' quantum fluctuations and
hamper the propagation of other triplets, a singly occupied
rung will do the same. For the triplets,
this effect has led to a quite dramatic
increase of the `energy of formation'
of the triplet. However, the corresponding renormalization
of the on-site energy for the hole-like Fermions is most likely
independent of the $k_y$ of the hole, and
hence can be absorbed into an overall
constant shift of the spectral function.
Since, the hole will not be able to propagate by two lattice
sites by coupling to a quantum fluctuation, so that
we expect that unlike the case of spin excitations
there will be no `dynamically generated' next-nearest neighbor hopping.\\
Performing the
Fourier and Bogoliubov transform we finally arrive at the
following Hamiltonian
\begin{eqnarray}
H &=& \sum_{\bbox{k},\sigma} 
\epsilon_{\bbox{k}}
a_{\bbox{k},\sigma}^\dagger a_{\bbox{k},\sigma}^{}
+ \sum_{\bbox{q}} \omega_{\bbox{q}} 
\bbox{\gamma}_{\bbox{q}}^\dagger \bbox{\gamma}_{\bbox{q}}^{}
\nonumber \\
&+& \frac{1}{\sqrt{N}}
\sum_{\bbox{k},\bbox{q}}\;[\; m(\bbox{k},\bbox{q}) \;
\bbox{\gamma}_{\bbox{q}}^\dagger 
\cdot
\bbox{S}_{\bbox{k},\bbox{k}-\bbox{q}}  + H.c. \;]
\nonumber \\
\epsilon_{\bbox{k}} &=& 2t_{eff}\; \cos(k_x) - t_\perp \cos(k_y),
\nonumber \\
m(\bbox{k},\bbox{q}) &=&
\delta_{q_y,\pi}\; [\; (2t \cos(k_x-q_x) + J\cos(q_x)) u_{q_x}
\nonumber \\
&\;&\;\;\;\;\;\;
- (2t \cos(k_x) + J\cos(q_x)) v_{q_x}\;].
\end{eqnarray}
Thereby it is understood that $\omega_{\bbox{q}}$
has only one branch with $q_y$$=$$0$.
Formally, this is already very similar
to the Hamiltonian derived by
Schmidt-Rink {\em et al.}\cite{SchmidtRink} for
hole motion in a Heisenberg antiferromagnet; the only differences
are that we have spinful holes, a nonvanishing
dispersion $\epsilon_{\bbox{k}}$ already for the `bare holes', and
$3$ branches of spin excitations rather than $1$.
In short, the Hamiltonian is explicitely spin-rotation invariant,
as it has to be in a `spin liquid'.
Technically this does not make any difference and 
the equation for the self-energy $\Sigma(\bbox{k},\omega)$ reads now
\begin{equation}
\Sigma(\bbox{k},\omega) = \frac{3}{N}
\sum_{\bbox{q}} \frac{ | m(\bbox{k},\bbox{q}) |^2}
{ \omega - \epsilon_{\bbox{k}-\bbox{q}} - \omega_{\bbox{q}}
-\Sigma(\bbox{k}-\bbox{q},\omega)}.
\end{equation}
The self-consistency equation can be solved for relatively
large systems (we used $N=$$200$), and the self-energy be used to
calculate the hole-like photoemission spectrum
\begin{equation}
A(\bbox{k},\omega) = -\frac{1}{\pi} \Im
\frac{1}{-\omega - \epsilon_{\bbox{k}} - \Sigma(\bbox{k},\omega) + i0^+}.
\end{equation}
If we want to compare this spectrum
with Lanczos results\cite{Haas}, some care is necessary.
The reason is that the spectral function we have calculated above
is the one for the creation of a `bare' hole. 
When calculating the photoemission spectrum, i.e. the
spectral function of the operator $c_{\bbox{k},\sigma}$, there
exists the possibility that the annihilation operator
`hits' a rung in a triplet state. This will lead to
terms of the form $a_{\bbox{k}+\bbox{q},\sigma}^\dagger 
t_{\bbox{q}}$ in the photoemission operator,
i.e. the creation of the hole is accompanied
by the annihilation of a spin excitation.
If we want to have numerical results for
the bare hole spectral function, we therefore should
use the Fourier transform of the operator
\begin{eqnarray}
\tilde{c}_{n,k_y,\uparrow}
&=& a_{n,k_y,\downarrow}^\dagger s_{n,n+\hat{y}}
\nonumber \\
&=& \frac{1}{2}[\; (\;c_{n,\uparrow} n_{n+\hat{y},\downarrow}
-  c_{n,\downarrow} S_{n+\hat{y}}^-\;)
\nonumber \\
&\;&\;\;\;\;
+ e^{ik_y}(\;c_{n+\hat{y},\uparrow} n_{n,\downarrow}
-  c_{n+\hat{y},\downarrow} S_{n}^-\;)\;].
\end{eqnarray}
This operator replaces a rung-singlet by a singly occupied bond
with the proper $k_y$, and annihilates any triplet.
This may therefore be considered as a creation operator for
a `bare' hole. Then, Figure (\ref{fig7}) compares the 
`bare hole' spectral function obtained by the
self-consistent Born approximation and the Lanczos spectra of the
operator $\tilde{c}_{\bbox{k},\sigma}$. The agreement is obviously
quite good, although the self-consistent Born result
tends to produce 
`too coherent' spectra and does
not put enough weight 
\noindent
into incoherent continua.
This may be an indication 
that our renormalization
of the coupling matrix element $m(\bbox{k},\bbox{q})$
by $(1-n_b)$ is too strong - a larger coupling 
would presumably lead to more incoherent weight.
The key features, however, are essentially identical in both
spectra: the intense band in the $k_y$$=$$0$ sector,
which disperses slightly upwards from $k$$=$$0$ towards
$k$$=$$\pi/2$ and then more or less levels off; the
equally intense band in the $k_y=\pi$ sector, which
starts out at $k$$=$$\pi$, disperses towards lower energy
and quickly becomes `overdamped'. The SCB results
somewhat underestimate the bandwidth of the
$k_y$$=$$0$ quasiparticle band and put the $k_y$$=$$0$
at a somewhat too high energy. The former
could probably be remedied by
adjustment of the `bare hole hopping integral' $t_{eff}$ -
however, since there is no rigorous way to do so,
we decided not to do this. Apart from these relatively minor problems,
however, there is quite good agreement.\\
As a last remark we note that a comparison with the
`full' photoemission spectrum\cite{Haas} shows quite
substantial differences in the spectral weight
of some features. The additional processes where the
photoemission operator annihilates a quantum
fluctuation (in this case a triplet) thus are
quite important for a discussion
of the true photoemission lineshape (see also
Refs. \cite{Eder,Sushkov} for a discussion of this
issue in the 2D systems).
\section{Conclusion}
In summary, we have derived a simple theory of two-legged spin ladders
which reproduces a number of numerical results quite well.
Being of comparable simplicity as linear spin wave theory
for the planar Heisenberg antiferromagnet,
the theory nevertheless allows to make quantitative calculations
of physical quantities, which in all cases compare favourably
with the results of Lanczos diagonalization. 
Just as linear spin wave theory may be viewed as
constructing an effective Hamiltonian for
the pair creation and propagation of fluctuations
around the `Neel-vacuum', the present theory
may be viewed as an entirely analogous expansion around
the `RVB-vacuum': the Hamiltonians (\ref{eff}) and
(\ref{eff1}) may be thought of as describing the dynamics of
triplet-like fluctuations around the RVB vacuum, and to discuss
the hole dynamics we basically needed to describe the coupling
between these triplet-excitations and the doped hole.
The great simplification which made the calculation possible
was the special geometry of the $2$-legged ladder which
immediately suggested a unique RVB-vacuum around
which we could `expand' the fluctuating ground state.
We note that there actually exists another system with such a
unique RVB-vacuum, namely the Kondo lattice. For this system
an analogous procedure involving {\em Fermionic} fluctuations
is possible and leads to excellent results when compared to
Lanczos diagonalization\cite{oana}.\\
In a two dimensional $t-J$-like system, such a unique vacuum does 
in general not exist, unless one were to assume some spatial inhomogeneities
such as stripes\cite{Zaanen}. Rather, for a translationally invariant
state, the most natural vacuum would be a statistical
average over all possible dimer-coverings of the plane, where
each dimer corresponds to a singlet\cite{Sutherland}. The analogue
of the triplet-like
excitations in the ladder then would be states where one or several
singlet-dimers are substituted by a triplet, and this excited dimer
propagates through the system. In fact, up to an additional
statistical averaging to account for the probability of suitable
dimer configurations,
the calculation of matrix elements for the pair creation
and propagation of these excited dimers proceeds in an entirely analogous
way as for the ladder\cite{tobepub}. We note that such a picture for the
low energy excitations of the RVB state is almost mandatory in the
framework of the $SO(5)$ symmetric theory of cuprate superconductors
by Zhang\cite{zhang}: there, the $\pi$-operator, which actually 
accomplishes the
$SO(5)$ rotations of spin-excited, half-filled states
into doped, superconducting ground states\cite{ederhankezhang}
precisely converts excited dimers with momentum
$(\pi,\pi)$ into $d_{x^2-y^2}$ symmetric hole pairs
with momentum $(0,0)$. Then, the simplest
picture of the SO(5) rotation from the antiferromagnetic to
the superconducting state\cite{zhang} would be 
that a condensate of excited dimers
with momentum $(\pi,\pi)$ at half-filling (i.e. a
state with nonvanishing staggered magnetization)
is converted into a condensate of $d$-like hole pairs 
(i.e. a $d$-wave superconductor) in the
doped case. A description of the spin excitations in $2D$ in terms of
Boson-like excited dimers as for the $t-J$ ladder thus
may be a very natural starting point for this promising scenario.\\
Instructive discussions with Dr O. P. Sushkov and
Dr E. Arrigoni are most gratefully acknowledged. 
\begin{table}
\narrowtext
\caption{The Fourier coefficients of the renormalized
energy $\tilde{\epsilon}_k$ and
pairing amplitude $\tilde{\Delta}_k$, extrapolated to
infinite length of the ladder.}
\begin{tabular}{ c c c c }
$\;$ & 0 & 1 & 2\\
\hline
$\tilde{\epsilon}_k$ & 1.77 & 0.21 & -0.32 \\
$\tilde{\Delta}_k$   & 0.07 & 1.03 &-0.17
\end{tabular}
\label{tab1}
\end{table}

 
\end{multicols}
\end{document}